# Drop Impact on Two-Tier Monostable Superrepellent Surfaces

Songlin Shi[1,2], Cunjing Lv[1,2,*], Quanshui Zheng[1,2,3*]

[1]Department of Engineering Mechanics, Tsinghua University, 100084 Beijing, China

[2]Center for Nano and Micro Mechanics, Tsinghua University, 100084 Beijing, China

[3]State Key Laboratory of Tribology, Tsinghua University, 100084 Beijing, China

**Abstract**

Superrepellency is a favorable non-wetting situation featured by a dramatically reduced solid/liquid contact region with extremely low adhesion. However, drop impact often brings out a notable extension of the contact region associated with rather enhanced water affinity, such renders irreversible breakdowns of superhydrophobicity. Here, we report an alternative outcome, a repeated Cassie-Wenzel-Cassie (CWC) wetting state transition in the microscale occurs when a drop impacts a two-tier superhydrophobic surface, which exhibits a striking contrast to the conventional perspective. Influences of material parameters on the impact dynamics are quantified. We demonstrate that self-cleaning and dropwise condensation significantly benefit from this outcome – dirt particles or small droplets in deep textures can be taken away through the transition. The results reported in this study allows us to promote the strategy to design functional superrepellency materials.

[*] To whom correspondence should be addressed. Email: cunjinglv@tsinghua.edu.cn, zhengqs@tsinghua.edu.cn

# Introduction

Superhydrophobicity is realized by a combination of chemical hydrophobicity and micro- and/or nanotextures.[1-4] Materials with superhydrophobicity exhibit ultralow adhesion with water drops, due to the largely reduced solid-liquid contact region with air entrapped at the bottom. As a result, spectacular properties such as self-cleaning, anti-fouling and anti-icing are attainable.[5-16] Superhydrophobic surfaces (SHSs) are usually fabricated by employing various kinds of textures, such as micro/nano-sized pillars,[6,17,18] nanowires,[19-21] particles,[22-24] etc. However, transitions from the Cassie to the Wenzel wetting states generally happen[25-27] on these surfaces, suffering from unavoidable disturbance from natural environments such as impact of raindrops or evaporation. During the transition, liquid impales the textures, the materials lose the superhydrophobicity and the peculiar functions as well. In the past decades, to pursue practical applications, researchers have spent extensive studies on investigating the mechanism of wetting transition and fabricating robust superhydrophobic materials.[5,6,13,24,28-30] Achieving a reversible transition from wetting to non-wetting states is one of the most important directions to maintain the spectacular properties of superhydrophobicity. For a long time, people have believed that energy barrier exists between the Wenzel and Cassie wetting states, a reverse process (from the Wenzel to Cassie wetting states) cannot be achieved spontaneously[31-35] expect for employing external assistances such as condensation,[36] mechanical vibration,[37] heating and evaporation,[38,39] etc.

Very recently, the existence of monostability was reported,[40] which verifies that spontaneous transition from the Wenzel to Cassie wetting states can happen: although it was attained on one-tier microstructured surfaces for mercury, the monostability was only attained on two-tier nano-/microstructured surfaces if the liquid is water. However, understandings obtained in this study is based on an investigation of quasi-static processes. From a practical point of view, water drop impact is a more ubiquitous natural phenomenon. When a drop impacts the SHS, it first spreads and then retreats like a spring.[41-43] If the drop is able to rebound very quickly and leaves materials completely without any residual, it can be very helpful for realizing self-cleaning and anti-icing. For instance, decreasing the impact time of drops on textures becomes one important direction and is promising for anti-icing.[44-47] However, for common bistable SHSs, additional inertia would make water prone to wet the textures, leading to an irreversible transition from the Cassie to Wenzel wetting states. The Wenzel to Cassie wetting states transition is detrimental for nonwetting functionalities and should be avoided. A better realization of self-cleaning requires that dirt particles could be taken away with water not only from the surface but also from the inner textures. Moreover, when drops with a normal temperature impacts a cold SHS, a rapid decrease of the temperature of the liquid (due to the effective thermal conductivity between water and solid thorough the contact interface) would enhance adhesion, which makes the wetting behavior very different from the quasi-static monostability state.[40] In these contexts, wetting behaviors of drop impact on the monostable superrepellent materials remains largely unknown and needs to be understood. Most of the previous works focused on the transition from the Cassie to Wenzel wetting states when a drop impacts the hydrophobic/superhydrophobic textures,[26,27,48] a reverse transition process is rarely reported.

Here, in striking contrast to the conventional perspectives, we report a repeated wetting transition between the Wenzel and Cassie wetting states in the microscale, when a drop impacts the two-tier monostable SHSs. The spreading and retracting behaviors, as well as scalings obeyed, are figured out, which are extremely helpful for materials design to better realize water repellency. The Cassie-Wenzel-Cassie (CWC) wetting state transition in the microscale creates an opportunity that enables liquid get impaled the deep textures to take dirt particles or small condensed droplets away. By means of the repeated transition process, liquid could more intensively contact the base of the substrates, which in turn significantly enhanced the efficiency of removal ability in self-cleaning and dropwise condensation. We hope this work will offer new perspectives on materials development in a wide range of applications from self-cleaning, to high-efficiency condensation, and to anti-icing.

## Results

**Wetting state transition.**

To distinguish the unique features of this study from the conventional phenomena, we first make comparisons between drop impacts on bistable and monostable SHSs. As shown in Figure 1A, silicon wafers patterned with a square array of micropillars manufactured by photoetching are employed, with pillar width $a = 25$ μm, pillar height $h = 100$ μm and spacing (side to side) $b = 75$ μm. In our experiments, we make the pillar-structured substrates having the properties of bistability ($\theta_r = 145 \pm 3°$) and monostability ($\theta_r = 159 \pm 2°$), respectively, through two different treatments (more details are given in Methods and Figure S1). $\theta_r$ stands for the receding contact angle achieved on these substrates. To define the geometrical parameters, as shown in Figure 1B, we give a frame showing an instant during the drop impact (for convenience on the monostable SHS, but also applicable to the bistable one). We use $D_M$, $D_S$ and $D_W$ to denote the instantaneous diameters of the main drop, water suspended on the micropillars and Wenzel contact region, respectively. Figure 1C,D demonstrates the impact processes of drops on the bistable and monostable SHSs, in which the Weber number is We = $\rho U^2 D_0/\gamma$ = 29, denoting $\rho$, $U$ = 1.02 m/s, $D_0$ and $\gamma$ the mass density of water, impact velocity (i.e. the velocity when the drop touches the substrate), initial diameter and surface tension of the drop, respectively. From Figure 1C, it is seen that the drop firstly touches the micropillars (0 ms) and then spreads with a pancake shape to the maximum in diameter (2.1 ms), meanwhile water penetrates to the pillars and leads to the Wenzel state. Shortly after that, the contact region tends to retreat but the solid-liquid-vapor three-phase contact line pins. Finally, the Wenzel wetting state maintains (Movie S1).

However, as shown in Figure 1D, it is seen that after touching (0 ms), water is forced into the cavity of the micropillars in the contact region and forms a thin liquid layer, meanwhile the front edge of this layer spreads outward (0 - 0.7 ms) to a maximum. Then, the thin layer in the micropillars starts to shrink while the upper main drop is still spreading. Shortly after that (2.2 ms), water in the Wenzel contact region is emptied. Then (2.4 ms), the main drop starts to shrink. Finally, the drop entirely detaches from the substrate (7.9 ms) (more details are given in Figure S2, and see Movies S2&S3). Even though the drop penetrates the micropillars and exhibit a Wenzel wetting state in the microscale (i.e. micro-Wenzel), we have to emphasize that previous results suggest that the contact region between water and the nanostructures is still in a Cassie wetting state.[33] Interestingly, in the time span ranging from 0 ms to 5.2 ms, wetting state transitions from micro-Cassie to micro-Wenzel, and from micro-Wenzel to micro-Cassie, occurs repeatedly, namely "mCWC" transition in the following. These phenomena are quite unusual and remarkably different from Figure 1C and other common cases of monodirectional Cassie to Wenzel wetting transitions,[31,34,48-51] ending up with irreversible breakdowns of superhydrophobicity.

Detailed dynamic behaviors corresponding to Figure 1C,D are quantified in Figure 1E,F, respectively. As shown in Figure 1E, for drop impact on the bistable SHS, $D_M$ and $D_S$ firstly increase with time and then decrease, and almost reach a stable value (more than 20 ms due to oscillations). However, $D_W$ firstly increases with time, then reaches a plateau, and after that it decreases further. These behaviors caused by contact angle hysteresis are well-known.[43] Compared with Figure 1E, the dynamics on the monostable SHS as shown in Figure 1F exhibits a very different scenario. In this case, $D_M$ and $D_S$ firstly increase with time, and reaches the maximum around 2.4 ms. After that, $D_S$ decreases monotonously to zero, which indicates an entire detachment of the drop from the SHS. The unique feature is that $D_W$ firstly increases with time and reaches the maximum value around 1 ms, and then decreases until its value reaches zero at 2.2 ms, which corresponds to a complete dewetting. This dewetting time has the same order of the capillary time,[41] i.e. $\tau_0 \sim (\rho R_0^3/\gamma)^{1/2} \approx 3.7$ ms with $R_0 = D_0/2$. Noticeably, the first mCWC transition process happens in the time span when the main drop is still spreading. Around 2.8 ms, water in the main drop is pushed into the cavity again and then completely dewets around 3.3 ms. After that, water continues to get into the cavity of the micropillars and then get out. In fact, the mCWC transition occurs four times in Figure 1F, but the third time (3.6 – 3.8 ms) is too prompt to be observed obviously (see more details in Figure S2 in an enlarged section).

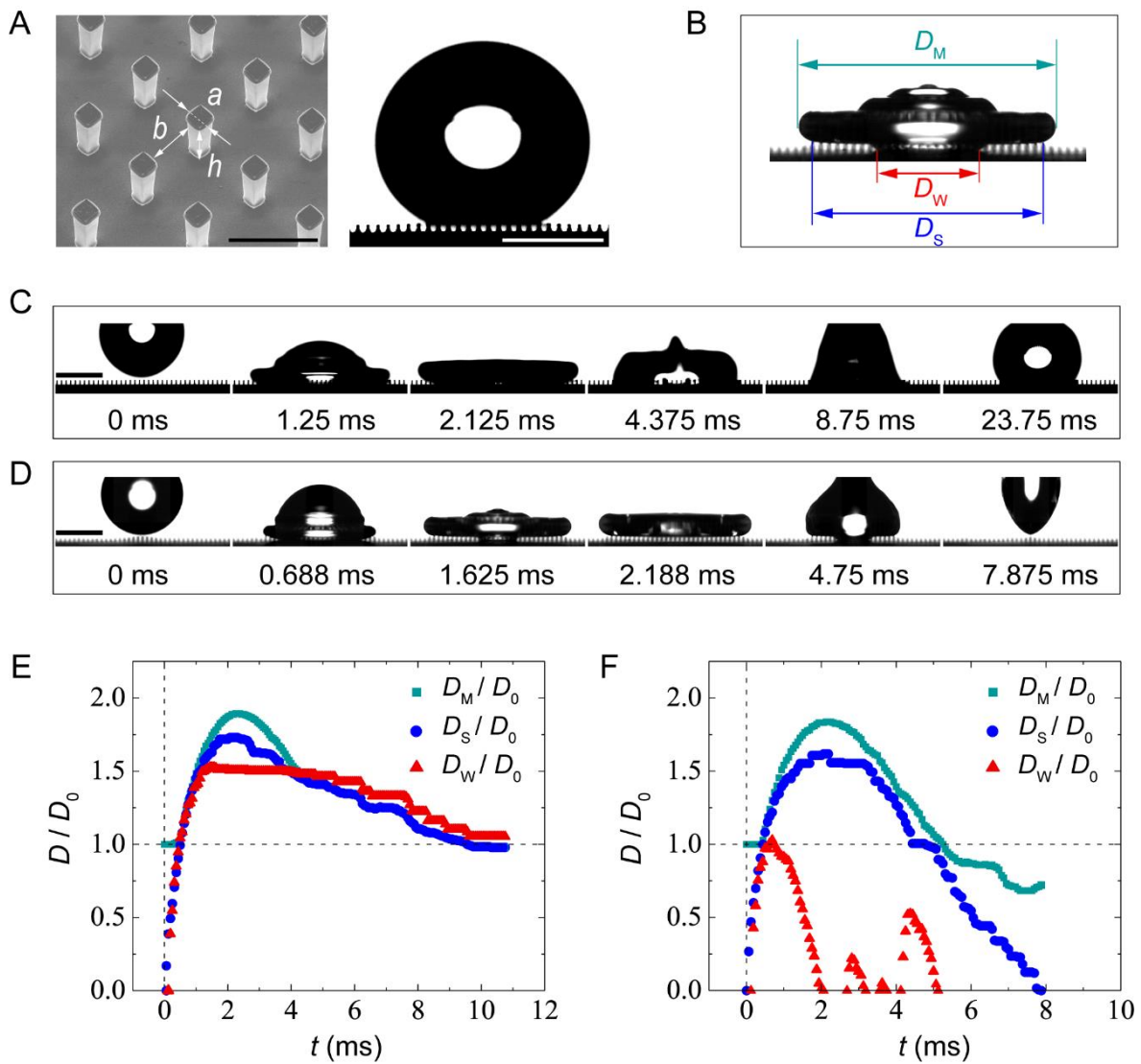

**Figure 1. Surface morphology and drop impact dynamics.** (**A**) Scanning electron microscopy (SEM) imaging of the Glaco-coated square-shaped micropillars fabricated on the silicon wafer. The width, spacing and height of the pillars are $a$ = 25 μm, $b$ = 75 μm and $h$ = 100 μm, respectively. The static wetting state of a 3.5 μL drop on this surface is shown on the right. The scale bars on the left and right represent 100 μm and 1mm, respectively. (**B**) $D_M$, $D_S$ and $D_W$ are defined in a snapshot captured on the monostable SHS, We = 29. (**C**)(**D**) Time-elapsed high-speed imaging of drop impacts on the bistable and monostable SHSs, respectively. The scale bars represent 1 mm. The evolutions of $D_M$, $D_S$ and $D_W$ corresponding to (**C**)(**D**) are quantified in (**E**)(**F**), respectively.

**Repeated and self-similar mCWC transition.**

To pursue a deeper understanding of the mCWC transition on the monostable SHSs, more details are elaborated. In Figure 2A, the first mCWC transition instants focusing on the contact region is shown in a high-speed motion capture under 16,000 frames per second (fps), and every three of the frames are aligned. The time span between the instants is 0.1875 ms (i.e. 3/16,000 ms). In Figure 2B, we give the instants corresponding to each maximum of $D_W$ of the four mCWC transitions (red triangles in Figure 1F). We reorganize the data in a dimensionless matter, i.e., $D_W/D_W^*$ vs $t_W/t_W^*$ in Figure 2C, denoting $t_W$ the time starting from the beginning of each mCWC transition, $D_W^*$ and $t_W^*$ the maxima of the Wenzel contact region and the corresponding Wenzel contact time in each CWC transition, respectively. The time error of the beginning of each mCWC is less than the time span of one frame, i.e. ~ 0.06 ms. Figure 2C shows that the data of each transition collapse into the same curve, which suggests the dynamics have a self-similar behavior. Furthermore, scaling law relations $D_W/D_W^* \sim (t_W/t_W^*)^{1/2}$ and $D_W/D_W^* \sim t_W/t_W^*$ are obtained in the spreading and recoiling stages, respectively (see Figure S3). A statistic of the parameter space of the micropillars (see Table S1) and Weber number (7 < We < 122) suggests that the repeated mCWC transition occurs at least twice (see Figure S4).

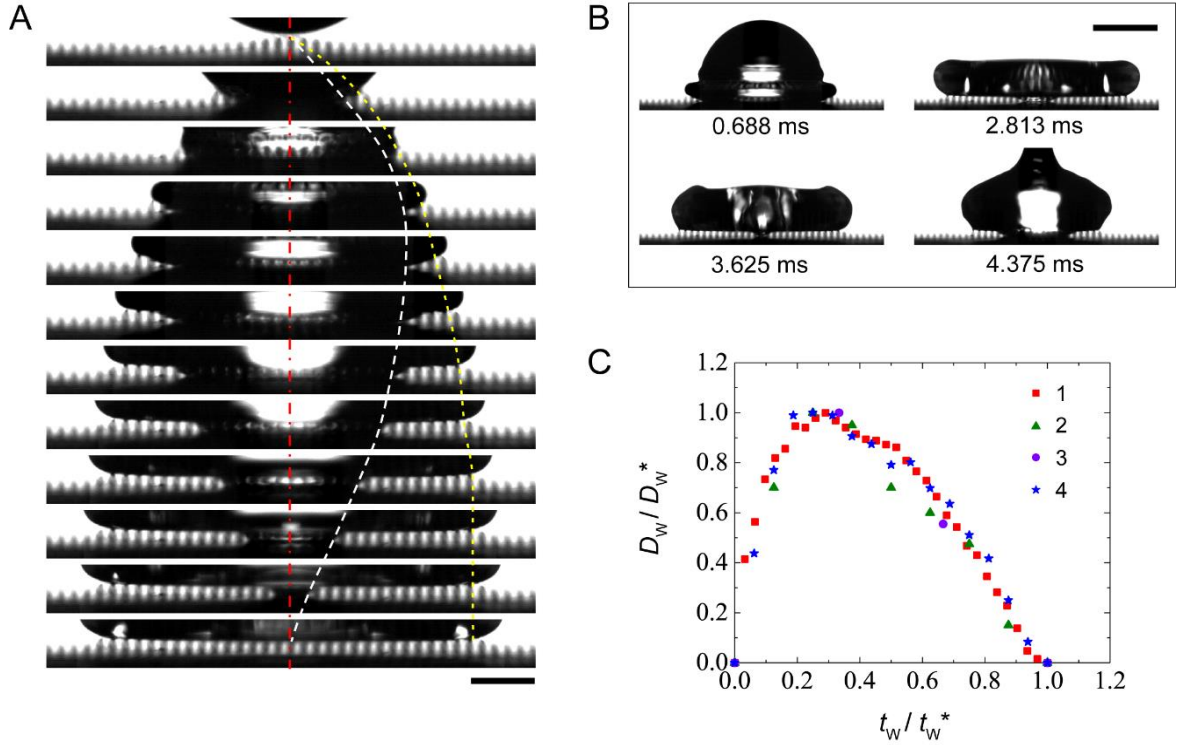

**Figure 2. Wenzel contact dynamics and self-similar transitions.** (**A**) Time-elapsed photographs showing details of the first mCWC process in the enlarged Wenzel contact region, We = 29. The time span between the instants is 0.1875 ms. The red dash-dotted line aligns the center of the drop, the dashed white and dotted yellow lines link the boundaries of $D_W$ and $D_S$, respectively. (**B**) Snapshots at the instant corresponding to the maximum of $D_W$ of each mCWC transition. (**C**) Normalized relationship between $D_W$ and $t_W$. Scaling performances with exponents 1/2 at the spreading and 1 at the recoiling stages are obtained, respectively (see Figure S3). Scale bars in (**A**) and (**B**) represent 0.5 mm and 1 mm, respectively.

**Mechanism for the evolvement of the Wenzel contact.**

We devote to understand the underlying mechanism accounting for the evolvement of the Wenzel contact and the mCWC transition. The unique features of the dynamic wetting behaviors on the monostable SHSs lie in the following aspects: (i) the process during the mCWC transition consists of two stages, the spreading and recoiling of the water layer confined in the micropillars; (ii) the mCWC transition is repeated during the impact process; (iii) when compare the recoiling with the classic Taylor-Culick problem[52,53] that usually the liquid layer has two same interfaces, the water layer in the micropillars shown in Figure 1D is constrained by the upper main drop and the lower substrate; (iv) furthermore, our experiments show a contact between the liquid and the bottom of the substrate, which is different from the pancake bouncing on quite deep micropillars that the liquid is still suspending.[45] These make our work distinguishable from the previous studies.

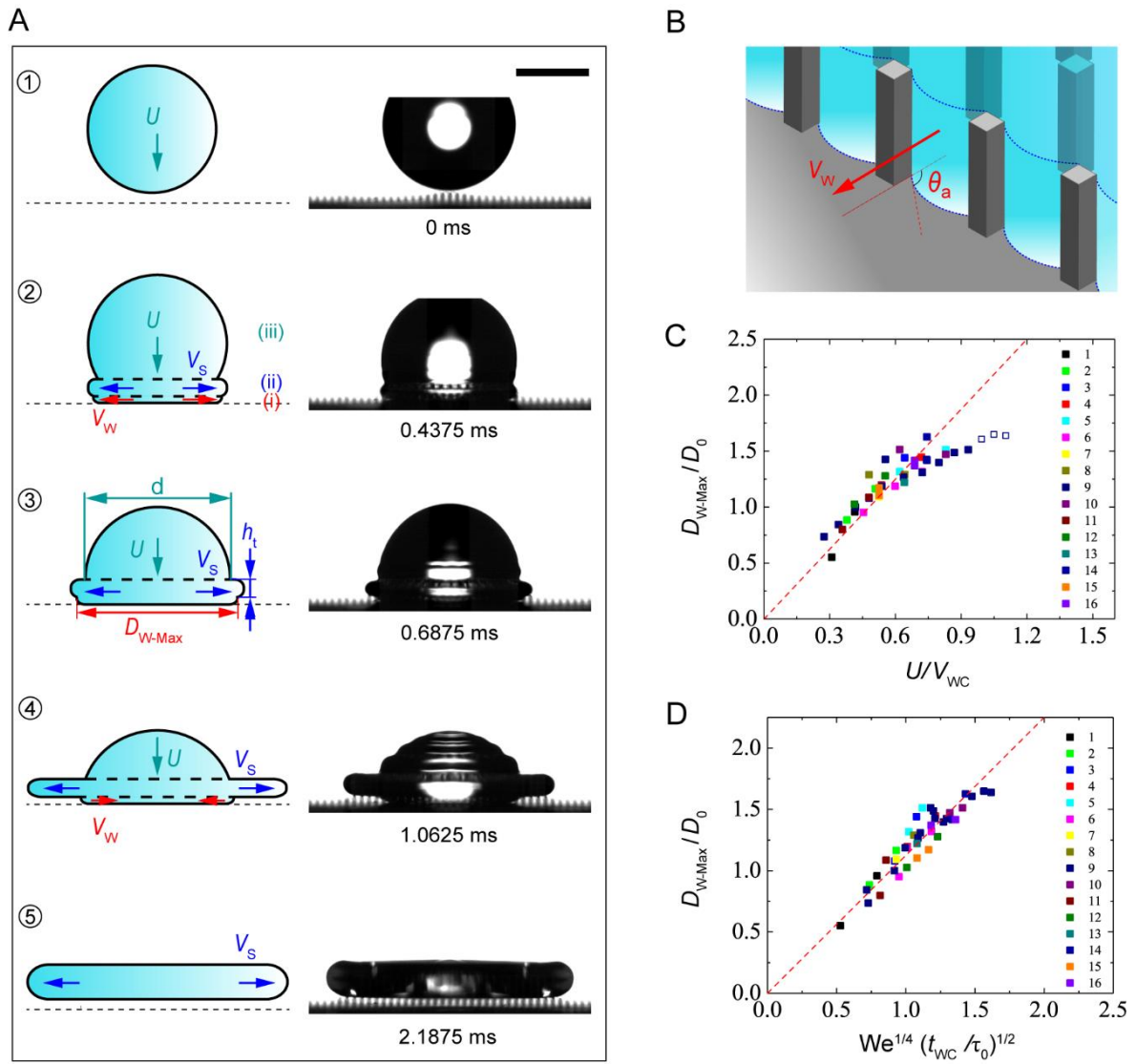

**Figure 3. Schematics and experiments illustrating the mCWC transition, and the scaling relations obeyed in the Wenzel spreading process.** (**A**) The drop touches the surface with an impact velocity $U$ ("circle 1"). After a while, three distinguished regions of the drop marked as (i), (ii) and (iii) are observed ("circle 2"), $V_W$ and $V_S$ denote the spreading velocities of regions (i) and (ii), respectively. In "circle 3", the Wenzel contact region reaches a stagnant point with a maximum diameter $D_{W-Max}$ while the lamella continuous to expand. $d$ denotes the diameter of the intersection of regions (ii) and (iii), and $h_t$ denotes the thickness of the lamella, i.e. region (ii). $V_W$ changes its direction in "circle 4" and the drop demonstrates a pancake shape in "circle 5". The scale bar represents 1 mm. (**B**) Schematic illustrating the advancing of the liquid-vaper interface of region (i), during which water is squeezed into the micropillars with $V_W$ at the front edge. $\theta_a$ denotes the advancing contact angle between water and the side wall of the pillars. (**C**) Comparison between experimental data (dots) and Eq. (3) (dashed line) with a fitting factor 2.083. (**D**) Comparison between experimental data (dots) and Eq. (4) (dashed line) with a fitting factor 1.125.

After the drop touches the substrate, it starts to spread. Previous studies suggest that the instantaneous diameter of the main drop obeys $D_S(t)/D_{M\text{-}Max} = [3Ut/(8D_0)]^{1/2}$,[54] denoting $t$ the time and $D_{M\text{-}Max}$ the maximum value of $D_M(t)$. Our experiments obey the rule very well (see Figure S5 & Section S1). For modeling, we give schematics in conjunction with experimental snapshots of the impact process in Figure 3A, and divide the drop into three parts as shown in the second frame of Figure 3A: (i) the Wenzel contact region $D_W(t)$, referring the liquid trapped into the micropillars; (ii) the lamella dimeter $D_S(t)$, referring the thin liquid lamella between the upper spherical cap and part (i); (iii) the upper spherical liquid cap.

From the third snapshot of Figure 3A, we can see that the Wenzel contact region $D_W(t)$ reaches a maximum $D_{W\text{-}Max}$, which distinguishes the stages of spreading and recoiling. During the spreading of $D_W(t)$, we observe that the moving of the front edges of part (i) and part (ii) are synchronized, i.e. $D_S(t) \approx D_W(t)$, and obey the following scaling relation $D_S(t)/D_{S\text{-}Max} \approx D_W(t)/D_{S\text{-}Max} \sim (t/t_{Max})^{1/2}$ in which $t_{Max}$ corresponds to the instant when $D_S(t)$ reaches $D_{S\text{-}Max}$.[54] We are interested in how $D_{W\text{-}Max}$ arrives, and we ascribe the reason to the existence of the micropillars. Whether the front edge of the Wenzel contact region could spread further or not depends on the strength of the inertia of the liquid compared with the capillary force, the latter is determined by the geometrical topology of the pillars. In the past, researchers quantified the critical pressure when a drop is able to penetrate an array of micropillars, in a manner of vertical loading.[25] For the sake of simplify, we suppose that the liquid in the Wenzel contact region go through the micropillars horizontally (as shown in Figure 3B), such, the pressure raised by the inertia $\rho V_W(t)^2/2$ has to overcome the pressure raised by the capillary force $P_c = -2\gamma \cos\theta_0 (a + h)/(bh)$ (see Figure S6 & Section S2), leading to a critical velocity of $V_{WC}$

$$V_{WC} = \sqrt{4(-\cos\theta_0)\frac{\gamma}{\rho}} \cdot \sqrt{\frac{a+h}{bh}}, \qquad (1)$$

We denote $\theta_0$ the contact angle on the surface without textures, $V_W(t)$ the instantaneous spreading velocity of the Wenzel contact region, i.e., region (i), as marked in Figure 3A ("circle 2").

Since the moving of the front edges of part (i) and part (ii) are synchronized as aforementioned, on the basis of $D_{S\text{-}max}/D_0 \sim \text{We}^{1/4}$,[55] we obtain (see Section S3)

$$V_W(t) \sim \text{We}^{1/4} \frac{D_0}{\tau_0} \cdot \frac{D_{S\text{-}Max}}{D_W(t)}, \qquad (2)$$

At the instant when $D_W(t)$ reaches $D_{W\text{-Max}}$ (Figure 3A), the inertia force is equal to the capillary force, i.e., $\rho V_W(t)^2/2 = \rho V_{WC}^2/2 \approx P_c$. On the basis of this fact, as well as Eqs. (1)-(2), we could obtain (see Section S3)

$$\frac{D_{W\text{-Max}}}{D_0} \sim \text{We}^{1/2} \frac{(D_0/\tau_0)}{V_{WC}} \sim \frac{U}{V_{WC}}, \tag{3}$$

The above relations suggest that the maximum diameter of the Wenzel contact region $D_{W\text{-Max}}$ is controlled by the geometrical parameters of the micropillars and the Weber number. In order to further understand Eq. (3), we carry out experiments in a large regime of the parameter space, i.e. $a \in [12.5, 150]$ μm, $b \in [37.5, 200]$ μm, $h \in [50, 330]$ μm (detailed parameters of the pillars are given in Table S1) and We $\in [7, 122]$. Figure 3C shows that most of the experimental data collapse into the same line, i.e. Eq. (3). Considering influences due to drop splashing may raise (Movie S4), Eq. (3) deviates from the experimental result under a high Weber number (We > 78, denoted using blue hollow dots as shown in Figure 3C). The dots with different colors are numbered, corresponding to samples given in Table S1. Moreover, the relationship between $D_{W\text{-Max}}$ and the Wenzel spreading time $t_{WC}$ (when the diameter of the Wenzel contact $D_W(t)$ reaches $D_{WC}$) could be obtained

$$\frac{D_{W\text{-Max}}}{D_0} \sim \text{We}^{1/4} \left(\frac{t_{WC}}{t_{\text{Max}}}\right)^{1/2} \sim \text{We}^{1/4} \left(\frac{t_{WC}}{\tau_0}\right)^{1/2}, \tag{4}$$

in which $t_{\text{Max}}$ is proportional to $\tau_0$, independent of the impact velocity.[41,44,45] Figure 3D suggests that the experimental data follows Eq. (4) very well. A more general impact scenario at different pillar spacing and Weber number is revealed in Figure S4. When fix the pillar width, pillar height and Weber number, the contact time $t_{\text{Max}}$ of the main drop is almost constant, while the contact time $t_W^*$ and the Wenzel contact region $D_{W\text{-max}}$ in the first mCWC transition increases with the pillar spacing. Moreover, for a specific SHS, $t_W^*$ and $D_{W\text{-max}}$ increase with We. These experimental results are consistent with the theoretical analyses.

Next, we study the recoiling process of the Wenzel contact region $D_W(t)$. As shown in "circle 4" associated with red arrows in Figure 3A, during the recoiling, the driving force exerted on the liquid-vapor interface is also resulted from the capillary force. Here we take a unit cell as the study object (see Section S3). The surface energy stored in such cell $\gamma\cos\theta_0[(a+b)^2 - a^2 + 4ah]$ transfers to the kinetic energy $mV_r^2/2$, which induces

$$V_r = \sqrt{-\frac{2\gamma\cos\theta_0}{\rho}\left(\frac{1}{h}+\frac{4a}{b^2+2ab}\right)}, \qquad (5)$$

denoting $m = \rho h[(a + b)^2 - a^2]$ the mass of water in the cell, $V_r$ the recoiling velocity of the liquid-vapor meniscus. The parameters are $\gamma = 0.072$ N/m, $\rho = 1000$ kg/m$^3$ and $\theta_0 = 160°$. Comparisons between Eq. (5) and the experimental results are given in Figure 4, in which the solid line denotes $V_r = V_{exp}$. Practically, it is challenging to capture the real-time performance of the three-phase contact line on the pillars so to accurately determine the direction of $V_r$. $V_r$ may have both horizontal and vertical components. In our estimation, it is simply postulated that $V_r$ is a result of the horizontal motion and which could be the reason accounting for the scattering of the data in Figure 4. Even though, the tendency of Eq. (5) and the experimental results are quite consistent with each other. Details of the flow field and the motion of the contact line will be further addressed in our future work.

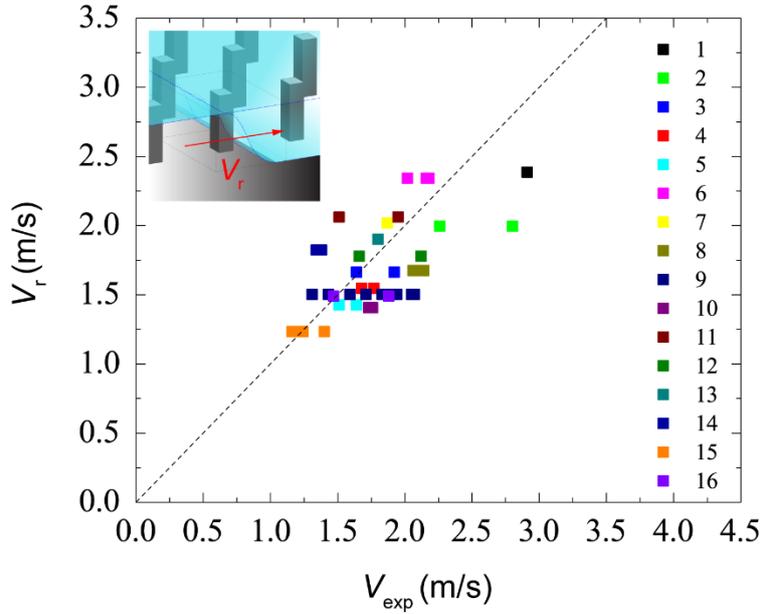

**Figure 4. Comparison between experimental results and theory of the recoiling of the Wenzel contact region.** The experimental data and theoretical model are demonstrated using dots and the dashed line, respectively, We > 20. The recoiling of the liquid-vapor interface in the micropillars are illustrated in the inset, denoting $V_r$ the recoiling velocity.

**Wetting transition enhanced self-cleaning and dropwise condensation.**

We expect the intriguing properties of the mCWC transition could find applications in many fields, especially in self-cleaning and dropwise condensation. It is well known that the famous self-cleaning effect refers to that dirt particles are picked up and taken away by water drops when they roll off the surface[2,56] (sketched on the left of Figure 5A). However, smaller particles would drop in the valley of the textures and accumulate, as a consequence, the SHSs becomes sticky and lose its self-cleaning ability. Here, we demonstrate that the CWC transitions could be used as a strategy to promote surface cleaning (sketched on the right of Figure 5A). As shown in Figure 5B, we carry out tests on the monostable micropillared SHS ($a$ = 25 μm, $b$ = 50 μm and $h$ = 100 μm) which is mounted on a platform with a 15° tilting angle and randomly deposited graphite powders having a mean size ~ 30 μm. Then, by employing a syringe, we first eject drops with a lower velocity (ranges from 0 – 0.4 m/s). After continuous drop impact on such contaminated SHS, the graphite powders on the top of the pillars are taken away, which is judged by the variation of the visualization of the surface. When compared with the clean surface (Figure 5B), the contamination remains in the valley of the textures even under more amount of drop impacting (see Movies S5). Notably, when employing drop impact under a higher velocity (0.9 – 1.8 m/s) to make water impale the pillars, the contaminations either on the top or in the valley of the textures are removed, and the surface becomes clean again under enough feeding (the lower slides of Figure 5C, see Movie S6).

Furthermore, realization of robust dropwise condensation is crucial for increasing heat transfer efficiency, anti-icing and water harvesting.[57-61] In most situations, the accumulated condensed droplets would lead to a sticky Wenzel wetting state (in both the micro and nanoscale).[15,62] How to make small droplets detach from the textures is challenging, but this topic bears a far-reaching potential for both fundamental and applied research. Here, through drop impact on the monostable SHSs, we demonstrate that small condensed droplets could be taken away so the possibility for further occurrence of the Wenzel wetting state can be suppressed. As shown in Figure 5D, we amount the SHS onto a cooling stage to perform condensation. When the surface is covered by a large number of condensed droplets, we perform drop impact in a low velocity (0 – 0.4 m/s), the condensed droplets on the top of the pillars are taken away, but the ones in the valley of the textures remain (the upper slides of Figure 5B, see Movie S7). Similar to the tests in Figure 5C, drop impact carried out in a higher velocity (0.9 – 1.8 m/s) could remove the condensed droplets either on the top or the valley of the pillars. With continuous feeding, more fresh regions on the sample are exposed and the dropwise condensation is able to be maintained (the lower slides of Figure 5D, see Movie S8 and Figure S7). We hope the mCWC transition has a potential to dramatically supress the filmwise condensation.

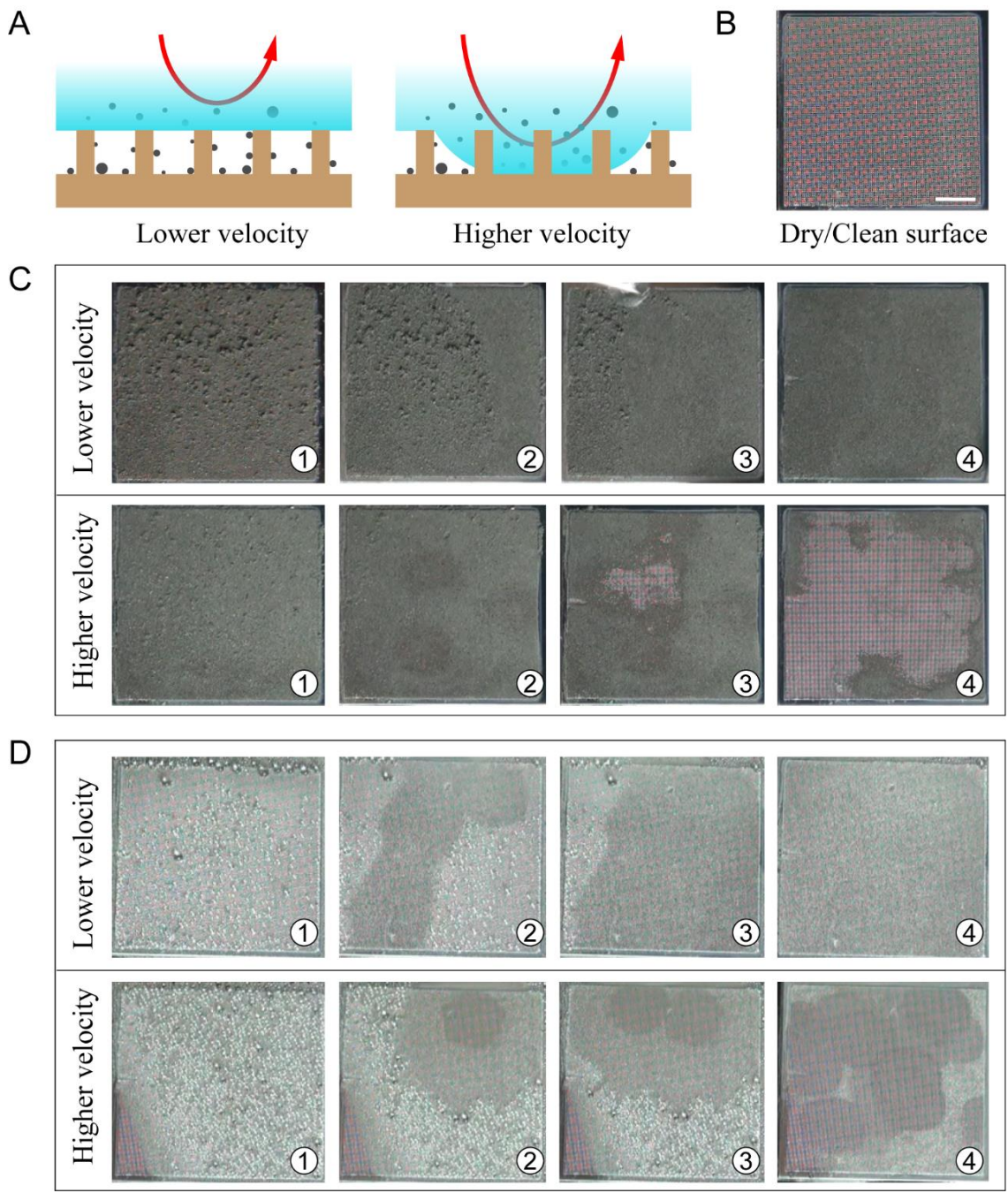

**Figure 5.** Wetting **transition enhanced self-cleaning and dropwise condensation.** (**A**) Schematics (not to scale) of the self-cleaning without and with the mCWC transition. (**B**) A piece of clean Glaco-coated monostable micropillared SHS. The scale bar represents 2 mm. The upper and lower slides of (**C**) and (**D**) respectively show the performances of continuous drop impact under lower (0 – 0.4 m/s) and higher (0.9 – 1.8 m/s) velocities on the graphite powder contaminated surfaces (**C**) and dropwise condensation (**D**) SHSs. The SHSs are amounted on a platform with a 15° tilting angle. The visualizations provide a judgement of the outcomes.

**Discussion**

Considering the first mCWC transition (0 – 2.2 ms, as shown in Figure 1F) is dominant among all the sub-mCWC transitions and it is the first step for the occurrences of the subsequent sub-mCWC transitions, it is worth being further explored, as the particular attention we have paid in the above. On the other hand, in the view of practical applications, owing the ability of sub-mCWC transitions is an advantage to enhance the efficiency of removal of dirt particles or small droplets in nature, because not only the amount of the contact area but also the intensity, are both largely promoted during the sub-mCWC transition. The investigations of the first mCWC transition in this work provide references for understanding of the other sub-mCWC transitions.

Decreasing the Wenzel contact region is desirable to avoid icing. On the basis of Eq. (5), we can see that a reduced Wenzel contact region benefits from decreasing the pillar height $h$. Meanwhile, decreasing $h$ is a good strategy to increase the mechanical stability of the pillars.[63] Eq. (5) also indicates that decreasing the value of $b$ or increasing the value of $a$ is helpful to reduce the Wenzel contact region. Considering practically the textures would be polluted from environments (e.g., micro/nano size fine particulate matters (PM 2.5), or ultrafine particles). Practically, getting into the gaps is a prerequisite for the liquid to take the dirt particles out of the deep textures. The typical velocity of rain drops is up to $U \approx 10$ m/s,[64] which leads to a spacing $b \sim \gamma/(\rho U^2) \sim 1$ μm for impalement transitions. This estimate of the spacing could be treated as a lower bound to design self-cleaning textures meanwhile having the ability to realize mCWC transitions under rain impingement. Moreover, our understandings are consistent with the fact that nature plants such as the lotus leaf has microscale papillae and nanoscale waxes.[1] The ability to resist contaminates benefits from the combination of these two-tier textures, the nanoscale structures guarantee a high receding angle to meet the criterion of monostability,[40] so the mCWC transition could take place at the microscale to realize self-cleaning, meanwhile the existence of the microscale textures largely avoid all the fragile nanoscale waxes being directly exposed to dirt contaminations. In this context, SHSs having two-tier textures is superior to these composed of nanostructure purely.

**Conclusions**

In summary, we report a novel phenomenon about drop impact on the monostable SHSs. At the contact region between the drop and the substrate, repeated mCWC wetting state transitions have been observed. It is suggested that the parameters of the structure of the micropillars play an important role to control the drop impact behaviors. We have proposed models, in which competitions of the inertia resulting from the impact and the capillary force resulting from the textures lead to the maximum Wenzel contact region and the corresponding spreading time, as well as the recoiling velocity during the transition. In the view of practical applications, we demonstrate that the efficiency of self-cleaning and dropwise condensation can largely benefit from the mCWC transitions on the monostable SHSs, due to the ability of removing tiny objects in deep textures. Our results could also help to understanding the necessity of the multiscale SHSs adopted by nature. These findings will shed new light on design of monostable super water-repellant materials. While our work only takes consideration of micro-pillared SHSs, more topologies such as micro-cone and micro-pored surfaces with various dimensions need to be further explored.

## Materials and Methods

**Preparation of micropillared bistable and monostable SHSs.**

In our experiments, micropillars with square-shaped cross-sections were fabricated on silicon wafers using photolithography. To produce the hydrophobicity, the substrate was then treated using a molecular vapor deposition technique by using FDTS (Perfluorodecyltrichlorosilane). On the FDTS-treated smooth surface, the intrinsic contact angle is measured to be $104 \pm 2°$, while the advancing contact angle (ACA) and the receding contact angle (RCA) $107 \pm 2°$ and $100 \pm 3°$, respectively. With micropillars (see No. 8, Table S1), the apparent contact angle is $152 \pm 3°$ with ACA $158 \pm 3°$ and RCA $145 \pm 2°$. To make the surface more water-repellent, silanized silica nanobeads with diameter 80 nm dispersed in isopropanol (Glaco, Soft99) was employed. Firstly, we immersed the substrate into the Glaco solution and then dried it naturally in the air. Then we put the sample in drying oven 10 min under temperature 150°C. The above steps were repeated three times. After that, the RCA and ACA of the Glaco-treated smooth surface increase to about $152 \pm 2°$ and $164 \pm 3°$, respectively, while on the Glaco-treated micropillared surface the RCA is about $159 \pm 2°$ and ACA is about $167 \pm 2°$. All the contact angles are measured by employing a commercial contact angle measurement (OCA20, Dataphysics, Germany). The wettability of the monostable SHSs are characterized by the spontaneous W2C transition,[40] i.e. $(1 - f)/(r - f) < -\cos\theta_r$, denoting $f$ the solid-liquid area fraction and $r$ the roughness factor.

**Characterization of the SHSs.**

Micropillars and nanotextures shown in Figure 1A and Figure S1A,B are characterized by using scanning electron microscopy (SEM) (Quanta FEG 450) and atomic force microscopy (AFM) (NTEGRA Aura, NT-MDT, Russia).

**Performance of the experiments.**

We use a syringe pump to control the volume of the drops and a homemade stage to control the releasing height. By this way, we are able to vary the value of the Weber number. Schematics of the setup is shown in Figure S1C. In our experiments, we use deionized (DI) water. The syringe pump injects with a very smooth speed 0.1 μL/s. A commercial high-speed camera (FASTCAM Mini UX 100, Photron, USA) with 16,000 fps is used for capturing the impact process. To test the enhanced capabilities of self-cleaning (on the graphite powder contaminated surfaces) and dropwise condensation, a colorful camera (HXR-MC58C, Sony) with 25 fps is employed for a better visualization.

**Data availability.**

All data supporting the findings of this study are available within the article and its Supplementary Information or from the corresponding author upon reasonable request.

**Supplementary Materials**

Section S1. Drop impact at low velocity

Section S2. Critical pressure accounting for horizontal moving of the liquid meniscus

Section S3. The deduction of $D_{W\text{-Max}}$ during the spreading process

Figure S1. Surfaces characterization and experimental setup.

Figure S2. Characteristic frames of the drop impact process.

Figure S3. Dynamics of the first mCWC transition.

Figure S4. Variations of the diameters with time.

Figure S5. Spreading and recoiling processes during the impact.

Figure S6. Schematic of the impalement of the meniscus.

Figure S7. Drop impact on the SHSs under condensation.

Table S1. Geometrical parameters of the microstructures and the corresponding critical receding contact angle $\theta_r$.

Movie Legends

Movie S1. Drop impact on the bistable SHS.

Movie S2. Drop impact on the monostable SHS.

Movie S3. Detailed processes of the first Cassie-Wenzel-Cassie (mCWC) transition.

Movie S4. Drop impact on the monostable SHS with higher We.

Movie S5. Self-cleaning under a lower impact velocity.

Movie S6. Self-cleaning under a higher impact velocity.

Movie S7. Enhancing condensation under a lower impact velocity.

Movie S8. Enhancing condensation under a higher impact velocity.

**Acknowledgements**

**General**: S.S. thanks Pengfei Hao and Dong Huang for their help for supporting samples and surface fabrications. S.S. thanks Yanshen Li for helpful discussions and Maosheng Chai for AFM testing support.

**Funding:** Q.Z. and C.L acknowledge the financial support from the National Natural Science Foundation of China (Grant No. 11632009, 11872227). C.L. acknowledges the financial support from Tsinghua University (Grant No. 53330100318).

**Author contributions:** S.S, C.L. and Q.S. conceived the project and wrote the paper. S.S, and C.L. analyzed the data. S.S. carried out experiments. C.L. and Q.S. supervised the work.

**Competing interests:** The authors declare no competing financial interest.

**Data and materials availability:** All data needed to evaluate the conclusions in the paper are present in the paper and/or the Supplementary Materials. Additional data related to this paper may be requested from the authors.